# Transition States Energies from Machine Learning: An Application to Reverse Water-Gas Shift on Single-Atom Alloys

Raffaele Cheula[1] and Mie Andersen[1,2]*

[1]Center for Interstellar Catalysis, Department of Physics and Astronomy, Aarhus University, 8000 Aarhus C, Denmark

[2]Aarhus Institute of Advanced Studies, Aarhus University, 8000 Aarhus C, Denmark

*mie@phys.au.dk

## Abstract

Obtaining accurate transition state (TS) energies is a bottleneck in computational screening of complex materials and reaction networks due to the high cost of TS search methods and first-principles methods such as density functional theory (DFT). Here we propose a machine learning (ML) model for predicting TS energies based on Gaussian process regression with the Wasserstein Weisfeiler-Lehman graph kernel (WWL-GPR). Applying the model to predict adsorption and TS energies for the reverse water-gas shift (RWGS) reaction on single-atom alloy (SAA) catalysts, we show that it can significantly improve the accuracy compared to traditional approaches based on scaling relations or ML models without a graph representation. Further benefitting from the low cost of model training, we train an ensemble of WWL-GPR models to obtain uncertainties through subsampling of the training data and show how these uncertainties propagate to turnover frequency (TOF) predictions through the construction of an ensemble of microkinetic models. Comparing the errors in model-based vs DFT-based TOF predictions, we show that the WWL-GPR model reduces errors by almost an order of magnitude compared to scaling relations. This demonstrates the critical impact of accurate energy predictions on catalytic activity estimation. Finally, we apply our model to screen new materials, identifying promising catalysts for RWGS. This work highlights the power of combining advanced ML techniques with DFT and microkinetic modeling for screening catalysts for complex reactions like RWGS, providing a robust framework for future catalyst design.

## Introduction

Heterogeneous catalysis plays a fundamental role in industrial chemistry by improving the efficiency of catalytic processes and reducing energy consumption.[1] The activity and selectivity of catalyst materials are linked to the energy of transition states (TSs), the highest points in the energy paths of the elementary steps, of which typically one or two are the bottlenecks along the reaction coordinate.[2] A detailed understanding of TSs is therefore crucial for the rational design of catalysts. However, the identification of TSs using first-principles calculations is challenging due to the complexity and unknown topology of the potential energy surfaces.[3]

Brønsted-Evans-Polanyi (BEP) relations are simple models used to estimate TS energies, assuming a linear correlation between activation energies and reaction enthalpies for a given elementary step.[4,5] These relations are particularly advantageous to reduce computational effort, as adsorption energies are much more easily calculated than TSs.[6] BEP relations are often combined with thermochemical scaling relations (TSRs), which are used to calculate adsorption energies from given energies of similar adsorbates (e.g., smaller molecules bonded to the surface with the same atomic species).[7–9] This allows to reduce the number of DFT calculations required to estimate the energetics of the reaction mechanism on an unknown material, based on a set of calculations on similar materials. These linear models have been successfully applied to study reactivity trends in heterogeneous catalysis.[6,10,11] However, their predictive accuracy is limited, and they often do not hold when the geometry of the TSs is different from one material to another.[12]

Machine learning (ML) has emerged as a powerful tool to address the challenges in the modeling of heterogeneous catalysis at the atomic level, offering flexible and accurate models to capture non-linear relationships in catalytic systems. Development of architectures and data sets for training ML interatomic potentials (MLIPs) has seen enormous progress in recent years, but it remains challenging to train MLIPs with sufficient accuracy for many different materials – as desired in screening studies – and for estimating TS energies of many different reaction steps.[13–15] Alternatively, ML can be applied to identify descriptors or simpler predictive models,[16–22] offering models that are typically more complex and accurate than previous descriptor-based approaches such as the d-band model[23,24] or linear scaling relations, but less complex and more physics-inspired and data-efficient than MLIPs. Instead of relying solely on features derived from the geometric structure, as done for MLIPs, these models exploit a variety of features such as electronic features related to the projected density of states (e.g., d-band center), electronegativity or surface work function or structural features such as coordination numbers. The prediction of adsorption energies from such models, in connection with scaling relations such as BEP relations for predicting TSs, has already been successfully applied to catalyst screening.[25,26]

To move beyond the limitations of linear scaling relations, it is, however, desirable to develop ML models for predicting not only adsorption energies but also TS energies.[27–30] It has been shown that BEP relations do not hold for more complex classes of materials such as single-atom alloys (SAAs).[31–33] These systems, composed of isolated metal atoms dispersed in a host metal matrix, can exhibit remarkably good activity and selectivity, especially for hydrogenation reactions.[34–38] While their ability to break scaling relations makes them extremely interesting for exceeding the maximal attainable activity or selectivity predicted by scaling relations and volcano plots, it also makes it challenging to develop reliable and efficient computational screening workflows.

Here, we extend a previously developed graph-based ML model for predicting adsorption energies of complex adsorbates based on the Wasserstein Weisfeiler-Lehman graph kernel and Gaussian Process Regression (WWL-GPR)[21,39] to now also allow for prediction of TS energies. We apply the methodology to a wide range of SAAs and a complex reaction network describing several possible reaction paths for the reverse water-gas shift (RWGS) reaction, i.e., the hydrogenation of $CO_2$ to CO and $H_2O$. Based on a DFT-calculated dataset comprising 1448 adsorbate and 650 TS structures on (111) and (100) facets of 6 pure metals and 12 SAAs, we find that WWL-GPR significantly improves the accuracy of both adsorption and TS energy predictions compared to linear scaling relations and ML models without graph representation. Applying the energy predictions in microkinetic modeling, we further show that the improved accuracy leads to a reduction of the errors in model-based versus DFT-based predicted turn-over-frequencies (TOFs) of almost one order of magnitude compared to the use of linear scaling relations. Finally, we apply our methodology to screen other SAAs and find several systems that may offer advantages as catalysts for RWGS compared to pure metals such as Ni and Rh, which, although active, suffer from low selectivity and/or susceptibility to deactivation by coking, especially at lower temperatures, due to their ability to further reduce CO to $CH_x$ species.[40,41]

## Methods

### DFT calculations

DFT calculations are performed with the Quantum Espresso[42–44] suite of codes using the BEEF-vdW exchange-correlation functional,[45] pseudopotentials from the SSSP library,[46] and a plane wave basis set. The plane wave and electronic density cut-off energies are set to 40 Ry (~544 eV) and 320 Ry (~4354 eV), respectively. The convergence threshold selected for the electronic energy of self-consistent field (SCF) steps is $10^{-6}$ Ry (~$1.36\cdot10^{-5}$ eV). For relaxation calculations, the convergence thresholds for the forces and the energy are set to $10^{-3}$ Ry/Bohr (~$2.57\cdot10^{-2}$ eV/Å), and

$10^{-4}$ Ry ($1.36 \cdot 10^{-3}$ eV), respectively. Adsorbates and TSs are calculated with 3×3 slabs (after a pre-relaxation on 2×2 slabs) with four metal layers, of which two are held fixed. A vacuum of 12 Å between periodic slabs is added in the z-direction. A Monkhorst-Pack mesh of 12×12×12 k-points is used for the face-centered-cubic (fcc) bulks and a proportional grid is used to map the Brillouin zone of the surface slabs. Spin-polarization is considered for systems containing Co, Ni, and Fe. The climbing-image nudged elastic band (CI-NEB)[3] method is used to identify transition states, with a 10 images path sampling and a final forces convergence threshold of 0.05 eV/Å. Vibration analyses are performed with the finite-differences method, as implemented in the Atomic Simulation Environment (ASE) library[47]. Displacements of +0.01 Å and −0.01 Å along the 3 cartesian coordinates are applied to the atoms of the adsorbates and TSs, and the vibrational frequencies are evaluated from the Hessian matrix. Gibbs free energies of reaction intermediates and TSs are calculated with the harmonic oscillator model. The vibrational contributions are calculated for the species on Rh(111) and attributed to the same species on the other materials.[48] The chemical potential of gas-phase molecules is evaluated from NASA coefficients based on experimental data[49] and the ideal gas approximation.

*Microkinetic calculations*

The catalytic activity of the materials under investigation is evaluated with mean-field microkinetic modeling. To enforce thermodynamic consistency in the microkinetic models, we apply a correction based on the analysis of the major sources of errors in the DFT-calculated gas molecule energies.[50,51] We identify the gas molecules that are responsible for the highest source of errors in the DFT calculations, and we correct their energy to reproduce experimental gas-phase reaction enthalpies. The resulting energy corrections are −0.114 eV for the CO molecule, +0.158 eV for $CO_2$, +0.581 eV for $O_2$, and +0.139 eV for $H_2$.

An ideal continuously stirred tank reactor (CSTR) model is used to calculate catalytic activities in terms of turnover frequency (TOF), i.e., moles of CO produced per moles of active sites per unit of time. A pure kinetic regime is assumed, neglecting mass transfer effects. The integration of the CSTR is done with the Cantera library.[52] The kinetic parameters of the microkinetic models are obtained with transition-state theory and kinetic theory of gases for non-activated steps. The mean-field approximation (MFA) is employed, and the resulting rate equations are solved at the steady state. Each surface is represented with a single generic site in the microkinetic model. The TOFs are calculated at conditions of 500 °C, 1 atm, a gas space velocity (GSV) of 100 $mol_{gas}/mol_{cat}/s$, and inlet gas molar fractions of 0.28 for $CO_2$ and $H_2$, 0.02 for CO and $H_2O$, and 0.40 for inert ($N_2$).

*Machine learning models*

We apply linear models (TSR and BEP), ensemble decision tree (EDT) models, and WWL-GPR to describe the formation energies of adsorbates and TSs. TSR and BEP relations are obtained with linear regressions and EDTs (random forest, XGBoost, and LightGBM) are applied within the Scikit-learn framework.[53]

The WWL-GPR model, developed for adsorbates (Figure 1.a), is herein extended to TS structures (Figure 1.b). In the WWL-GPR model, each atomic structure is associated with a graph. The nodes of the graph correspond to the atoms, while the edges are drawn in correspondence of the chemical bonds, typically obtained from atom distances. Here we draw a chemical bond if the distance between two atoms is lower than the sum of their covalent radii, plus 0.2 Å. The nodes are then decorated with a list of attributes, typically features of the clean surface (with no adsorbates) and the adsorbate molecule (in the gas phase). Here we use as node attributes (see Table S1): physical constants of the atoms (1-4), features of the projected density of states (PDOS) (5-12), features of the surface (13), geometrical features (14-15), features of the gas-phase molecules (16-17), features of the elementary step (18-20), and the SOAP[54–56] descriptor of the relaxed clean surface and molecules in the gas phase (21-90). In the WWL-GPR model, the graphs (after Weisfeiler-Lehman node information propagation in the graph and similarity measure through the Wasserstein distance metric[39]) are used to calculate adsorption energies with Gaussian Process Regression.

When applying the WWL-GPR model to adsorbates, we consider several possible graphs corresponding to pre-defined adsorption motifs (e.g., top, bridge, mono-dentate, bi-dentate) and the one with the lowest DFT- or model-predicted adsorption energy is then used in the microkinetic modeling. In the case of TS structures, the chemical bonds between the TS and the surface are not known in advance. Hence, in the TS graphs, we include the union of the edges that are in the initial state (IS) and final state (FS) structures. For node features, we use the same ones as for adsorbates (see Table S1), and we include also the energies of the initial state and the final state of the elementary step, where the latter is the only input used in BEP relations.

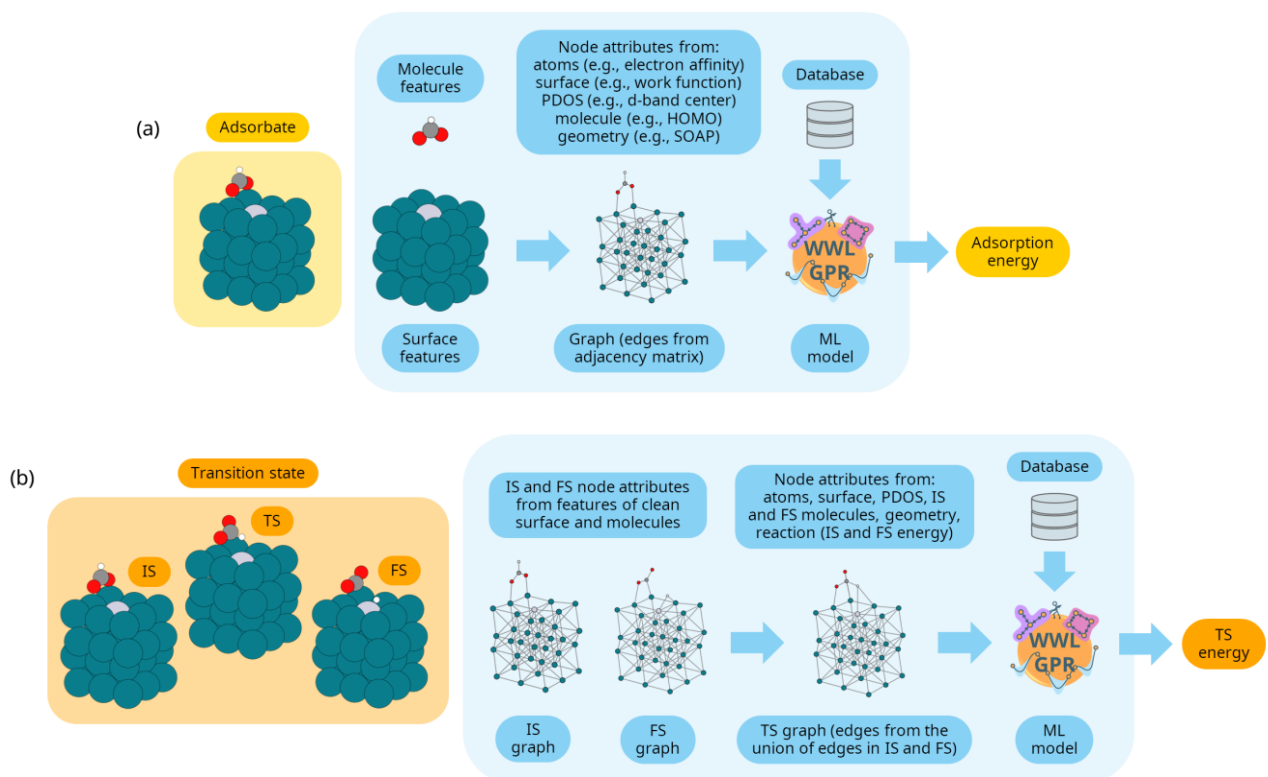

Figure 1: WWL-GPR model applied to (a) adsorbates and (b) TS structures.

In the application of EDTs, we use the same features used in the WWL-GPR model (Table S1) but averaged over the atoms of the surface site, i.e., the atoms bonded to the adsorbate (or TS). Hyperparameter optimization of the ML models is done with Bayesian search (with Scikit-optimize[57]) and randomized search cross-validations. During hyperparameter optimization, the test set is split into two subsets, which are iteratively used as validation and test sets (the validation set is used to select the best hyperparameters and the test set to calculate the errors).

**Results and discussion**

*DFT database*

We target the reaction mechanism of RWGS on two crystal facets, (100), and (111), of metals and SAA materials (illustrated in Figure 2.a). We perform DFT calculations on 6 metals (Rh, Pd, Co, Ni, Cu, and Au), and 12 SAAs (Rh+$Pt_1$, Pd+$Rh_1$, Co+$Pt_1$, Ni+$Ga_1$, Cu+$Zn_1$, Cu+$Pt_1$, Cu+$Rh_1$, Cu+$Ni_1$, Au+$Ag_1$, Au+$Pt_1$, Au+$Rh_1$, and Au+$Ni_1$). These SAAs are selected from the stable materials active for $H_2$ dissociation reported in a previous theoretical study.[22] We include in the study 10 reaction intermediates ($CO_2$*, CO*, O*, H*, OH*, $H_2O$*, COOH*, HCOO*, HCO*, COH*) and 14 elementary steps (reported in Figure 2.b). These include adsorption reactions ($CO_2$, $H_2O$, and CO adsorption, but not H₂ adsorption, are considered non-activated), the CO-O path (involving direct $CO_2$* dissociation) and $H_2O$ formation, the COO-H path (involving the $CO_2$* hydrogenation to COOH*, i.e., carboxyl), and the H-COO path (involving the $CO_2$* hydrogenation to HCOO*, i.e., formate). For mono-dentate reaction intermediates, we test the adsorption on all the unique symmetric sites of the surface structures, taking into account different types of top, bridge, and hollow sites determined by the geometry of the surface and the presence of the dopants. For bi-dentate reaction intermediates, we consider all the possible combinations of neighboring sites. For the identification of TSs for elementary steps with NEB calculations, we test different combinations of initial and final states, selected between the ones with the lowest energy. With this procedure, we obtain a dataset of 1448 unique adsorbate structures and 650 TS structures.

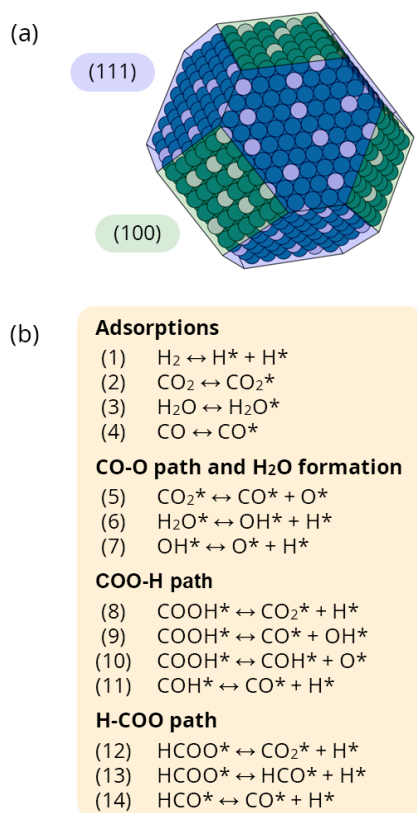

Figure 2: (a) Particle facets and (b) elementary steps of RWGS included in the study.

*Brønsted-Evans-Polanyi relations*

We analyze the dataset of TS energies with BEP relations, which correlate the activation energy ($E_{act}$) to the reaction enthalpy ($\Delta E_R$) of an elementary step on different materials. We report the plots in Figure 3 for (a) H₂ adsorption, (b-d) CO-O path and H₂O formation, (e-h) COO-H path, and (i-m) H-COO path. For some elementary steps, we observe an important effect of the facet geometry. For example, the $CO_2$* dissociation to CO* and O* (Figure 3.b) and the COOH* dissociation to COH* and O* (Figure 3.g) show an offset between the BEP relations obtained on (100) surfaces (red line) and (111) surfaces (orange line), in agreement with previous theoretical findings.[51,58] For other reactions, the presence of dopant atoms in SAA surfaces has an impact on the activation energies and they do not follow the BEP relations of the pure metals. Examples are reported in Figures 3.a, 3.c, and 3.m, for $H_2$ dissociation to 2H*, $H_2O$* dissociation to OH* and H*, and HCO* dissociation to CO* and H*, respectively. In some cases, SAAs have lower activation energies than predicted by the scaling relations fitted to the pure metals only. Finally, some elementary steps do not show good correlations at all between $E_{act}$ and $\Delta E_R$, e.g., HCOO* dissociation to $CO_2$* and H* (Figure 3.i) and COH* dissociation to CO* and H* (Figure 3.h).

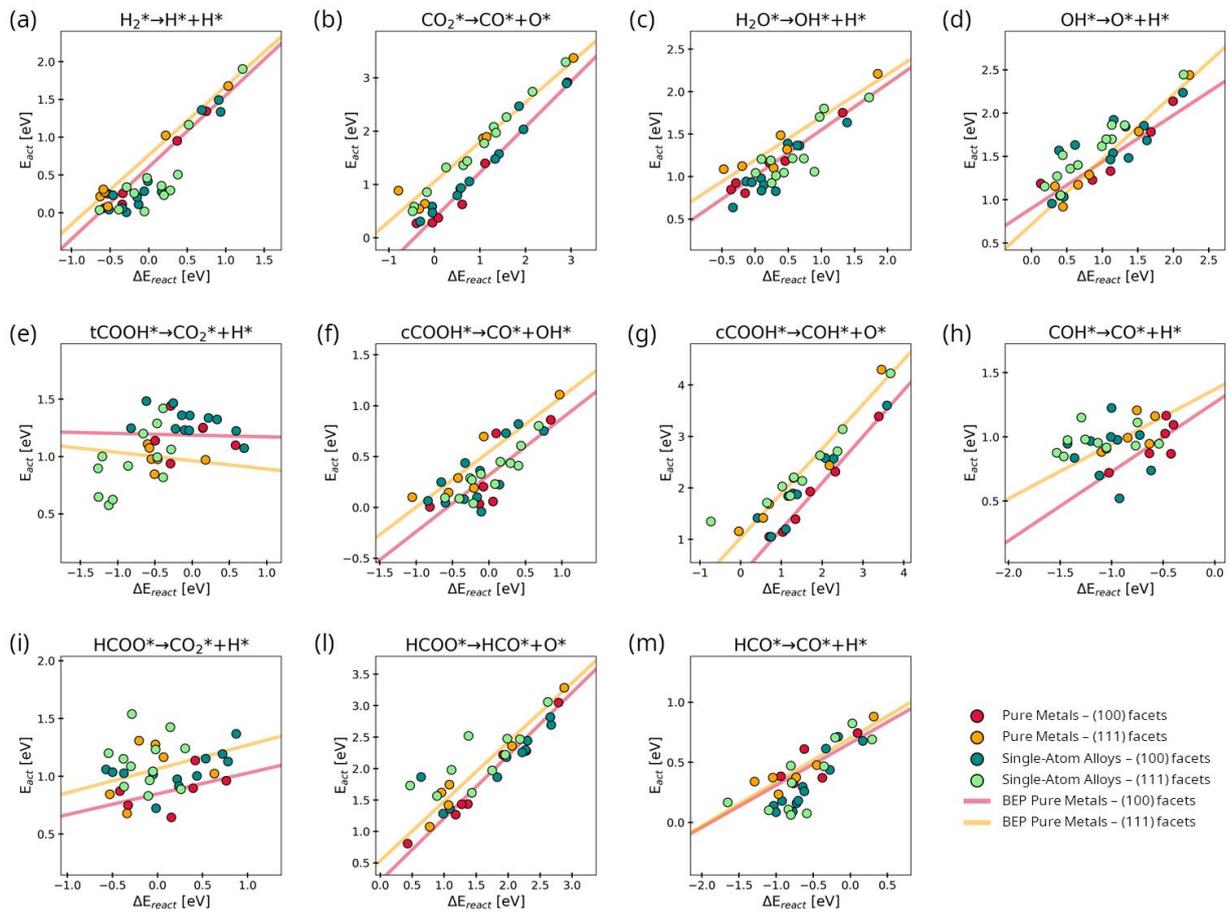

Figure 3: BEP relations for the activated RWGS steps included in this study.

*Performance of ML models for energy predictions*

We assess the performances of the different ML models under investigation (i.e., TSR, BEP, EDTs, and WWL-GPR), with stratified group K-fold cross-validation ensembles[59] (see Figure 4.a). The datasets are split into groups relative to different materials with a stratified group K-fold iterator, producing iteratively a train set (K-1 folds) and a test set (1 fold). Our DFT dataset contains 18 materials, so we use K = 6, resulting in folds containing 3 materials each. Stratification ensures that each adsorbate (or TS) species is well distributed in each fold. Then, to estimate the uncertainty in the predictions, we split the train set into K-1 folds using the same group K-fold iterator. We train separate ML models on each combination of K-2 folds (i.e. we sub-sample the train set), obtaining an ensemble of K-1 ML models. Each ML model in the ensemble is used to predict the energy of a given data point (adsorbate or TS) in the test set, generating a set of K-1 energy predictions for each data point. We use the standard deviation of these predictions as the uncertainty estimate and the mean as the predicted energy.

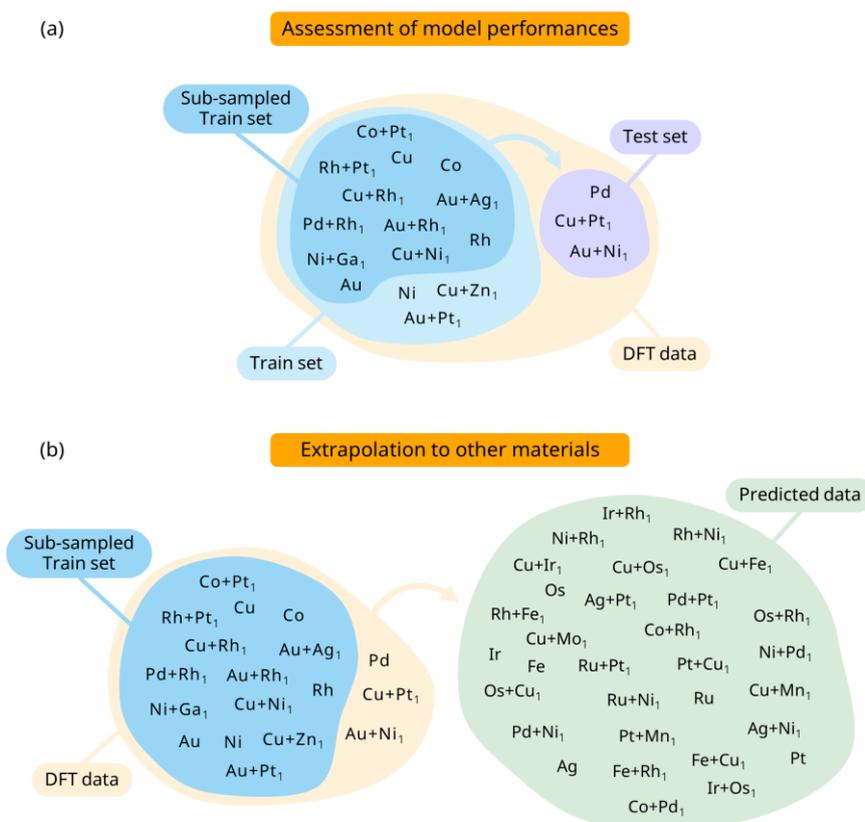

Figure 4: Scheme of data organization and repartition for (a) assessment of the model performances and (b) extrapolation to other materials.

In the TSR, we use CO*, H*, and O* as reference species for calculating the energies of other adsorbate species. When applying TSR, the energies of these reference species on the target surfaces must be known. In the ensemble cross-validation tests, we exclude these species from the test set and include them in the train sets to ensure a fair comparison between the ML models. The inclusion of the information on simple adsorbates on the target surfaces (which are otherwise not present in the train set) can significantly improve the performance of the ML models as well, as previously shown for WWL-GPR[21].

Figure 5 shows parity plots comparing model-predicted energies ($E_{model}$) with reference DFT values ($E_{DFT}$) for adsorbate (ads) and TS formation energies. Formation energies are calculated with respect to the clean surfaces and the atoms of the adsorbate (or TS) in the reference gas molecules (i.e., CO, $H_2$, and $H_2O$). The performance of six models is evaluated: TSR, LightGBM, and WWL-GPR for adsorption energies (panels a-c) and BEP relations, LightGBM, and WWL-GPR for TS energies (panels d-f). LightGBM is reported between the EDT models because it shows the best performances on both adsorbates and TS energies, as shown in Figure S1. The Mean Absolute Error (MAE) and Root Mean Square Error (RMSE) are reported to quantify accuracy.

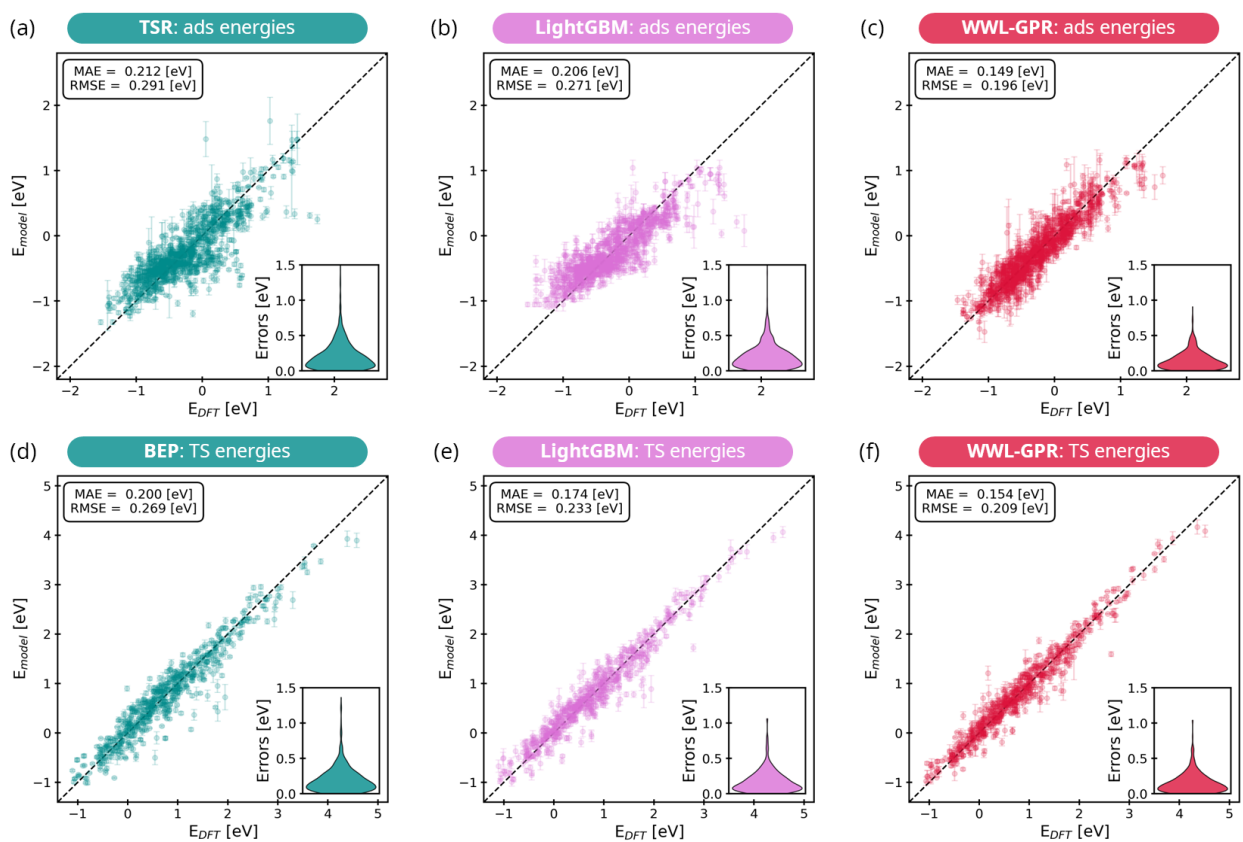

Figure 5: Parity plots comparing DFT-calculated adsorption (ads) and transition state (TS) formation energies with predictions from different ML models: (a, d) Linear models (TSR and BEP), (b, e) LightGBM, and (c, f) WWL-GPR. Insets show the error distributions.

For adsorption energies, TSR (panel a) exhibits the highest error (MAE = 0.212 eV, RMSE = 0.291 eV). LightGBM (panel b) marginally improves predictions (MAE = 0.206 eV, RMSE = 0.271 eV) by incorporating nonlinear relationships. WWL-GPR

(panel c) significantly enhances accuracy (MAE = 0.149 eV, RMSE = 0.196 eV), indicating that the graph representation is beneficial. We note here that the considered reaction network is dominated by rather simple adsorbates, compared to those considered in previous work employing WWL-GPR, and that the advantage of the graph representation is expected to become more pronounced for more complex adsorbates.[21] For TS energies, the BEP model (panel d) achieves MAE = 0.200 eV and RMSE = 0.269 eV, performing comparably to TSR for adsorption energies. LightGBM (panel e) improves predictions (MAE = 0.174 eV, RMSE = 0.233 eV). The best performance is observed also in this case with WWL-GPR (panel f), which achieves the lowest errors (MAE = 0.154 eV, RMSE = 0.209 eV). The error distributions, illustrated by the violin plots, show that WWL-GPR exhibits a more localized error distribution with fewer outliers, further confirming its robustness. Overall, these results suggest that WWL-GPR is the most effective approach for both adsorption and TS energy predictions, followed by LightGBM. The linear models (TSR and BEP), while widely used for energy estimates in catalysis, exhibit larger deviations, reinforcing the need for ML-based approaches for accurate energy predictions.

We further evaluate the WWL-GPR model on two distinct tasks, with the results displayed in Figure S2. The first task focuses on predicting the energies of adsorbates and TSs on surfaces included in the training set. For this, we apply a randomized stratified cross-validation method. The model shows lower errors compared to the group cross-validation results shown in Figure 5, with MAEs of 0.104 eV and 0.147 eV for adsorbates and TSs, respectively, and RMSEs of 0.160 eV and 0.197 eV. These errors are relatively low because the task is confined to in-domain surfaces, where the data is well-distributed across the various surface types. In the second task, we assess the model performance on surfaces containing chemical elements not present in the training set. For this, we use a group cross-validator that partitions the data based on the chemical elements of the surfaces. In agreement with previous observations[21], the model errors are higher in this out-of-domain scenario, although they are substantially reduced when data for CO*, H*, and O* species are included. In this case, the MAE and RMSE for adsorbates rise to 0.225 eV and 0.288 eV, respectively, while for TSs, they increase to 0.200 eV and 0.262 eV. This reflects the challenges of predicting energies for surfaces with chemical elements that are not represented in the original DFT training data.

*Performance of ML models for TOF predictions*

For application of the ML models in catalyst screening, it is desirable to keep the computational cost low by calculating only few, simple adsorbates on the surface to be screened. To estimate the error associated with this procedure, we now evaluate the performance of the ML models for predicting TS energies based on model-predicted adsorption energies (with the exception of the CO*, H*, and O* adsorption energies that are explicitly calculated and used to train the models). Specifically, we analyze two approaches: the combination of BEP and TSR models (Figure 6.a) and the WWL-GPR model for both adsorbate and TS energies (Figure 6.b). Compared to Figures 5.d and 5.f, the errors increase due to the propagation of errors from the adsorption energy models. The WWL-GPR model continues to exhibit the lowest errors (MAE = 0.207 eV, RMSE = 0.274 eV) compared to the BEP model (MAE = 0.274 eV, RMSE = 0.366 eV), confirming the better predictive power of the graph-based GPR model.

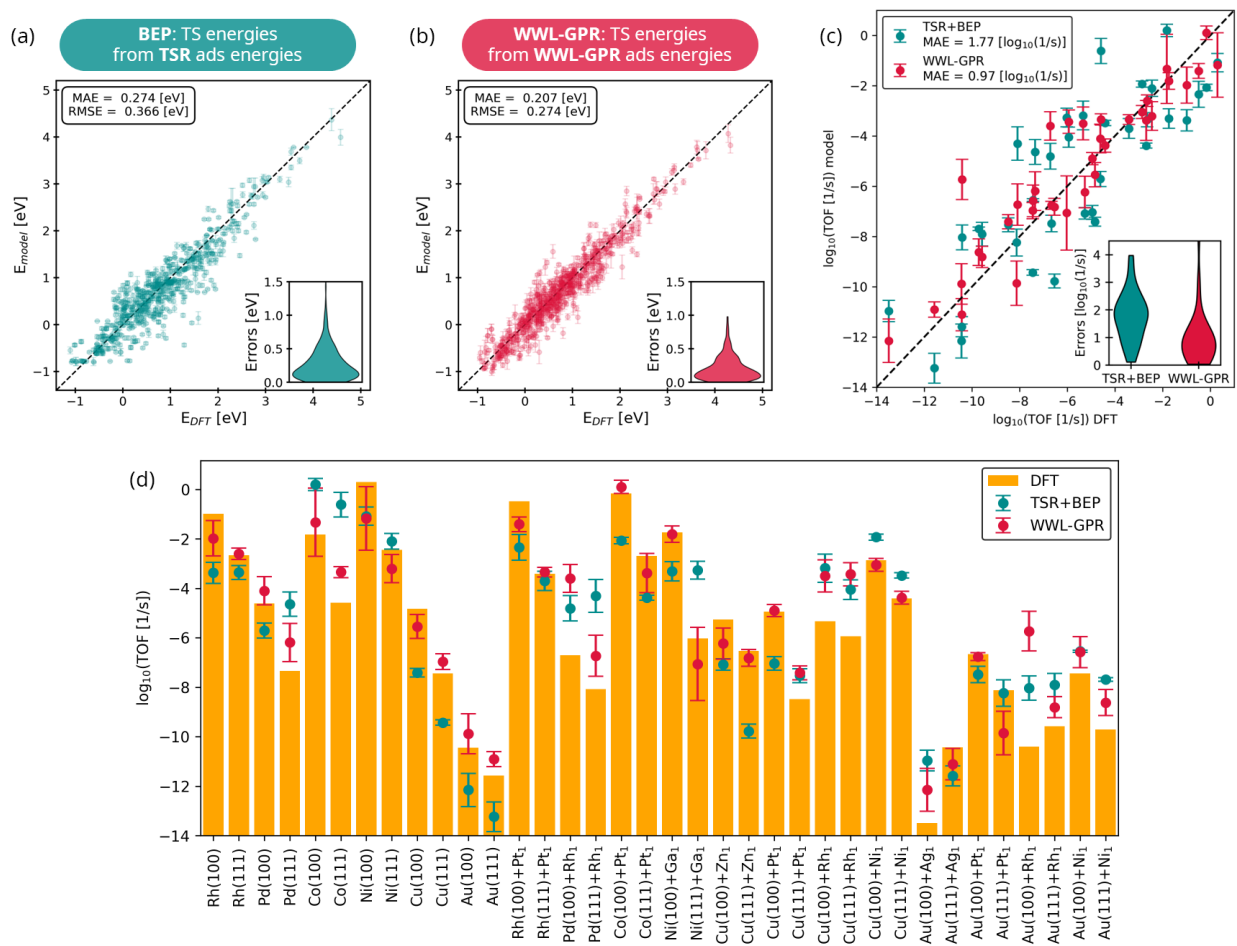

Figure 6: (a) Parity plot of TS energies calculated with BEP, using adsorption energies obtained with TSR, vs. DFT. (b) Parity plot of TS energies calculated with WWL-GPR, using adsorption energies obtained with WWL-GPR, vs. DFT. (c) Parity plot of calculated TOFs based on BEP+TSR and WWL-GPR, vs. DFT input. (d) Calculated TOF for each surface based on DFT, BEP+TSR, and WWL-GPR input.

Next, we assess error propagation to predicted RWGS TOFs. In Figure 6.d we show the TOF of each surface in the DFT dataset calculated with microkinetic models based on DFT energies (orange bars). To assess the accuracy of the ML models for predicting

TOFs, we show in the figure also the TOFs calculated using microkinetic models based on energies from the linear models (TSR+BEP, in green) and WWL-GPR (in red). By applying ensemble group cross-validations with train set subsampling (see Figure 4.a), we generate ensembles of ML models, yielding a list of predicted energies for each data point. For each surface, we construct an ensemble of microkinetic models, incorporating the energies from the corresponding ML model ensembles. The integration of these microkinetic models produces a distribution of TOF values, from which we take the average as the predicted TOF and the standard deviation as the associated uncertainty. Figure 6.c summarizes the errors in the TOF introduced when replacing DFT-calculated energies with energies from the ML models. We present the values and errors as the $\log_{10}$ of the TOF, which represents the order of magnitude of the TOF. The use of the linear models (TSR+BEP) results in a MAE of 1.77 orders of magnitude in TOF predictions, while the WWL-GPR model reduces the MAE to 0.97 orders of magnitude. This improvement can be attributed to the lower errors in adsorption energies (see Figures 5.a and 5.c) and TS energy predictions (see Figure 6.a and Figure 6.b), where the WWL-GPR model outperforms the linear models (TSR and BEP). The violin plot in Figure 6.c further illustrates that the WWL-GPR model yields a narrower error distribution compared to the linear models. These findings emphasize the importance of accurate adsorption and TS energy estimations in microkinetic modeling and demonstrate the advantage of WWL-GPR over linear scaling models in capturing complex energetic trends.

*Extrapolation to other materials*

After assessing the performances of the ML models for the prediction of energies and TOFs of RWGS, we use the best-performing model (WWL-GPR for both adsorbate and TS energies) to predict the TOF of a larger set of materials (6 metals and 26 SAAs). For this out-of-domain task (Figure 4.b), we split the entire DFT dataset into K folds, and we produce an ensemble of K ML models trained on each combination of K-1 folds, using again K = 6. We add to the sub-sampled train sets the data of CO*, H*, and O* adsorbed on the (100) and (111) surfaces of the new materials (478 adsorption energies), and we use the ML model to calculate the data of other reaction intermediates (3722 adsorption energies) and elementary steps (1472 TS energies). For each material, the contributions of the (100) and (111) crystal facets to the total TOF are weighted by their area fraction, calculated from a Wulff construction of the pure metals (see Figure S3).[60,61]

In Figure 7 we show the TOFs obtained from energies calculated with DFT (in orange) and with the WWL-GPR model (in red). Our findings are in good agreement with experimental work by Dai et al.[62], where $CO_2$ conversions were observed in the order of $Ni/CeO_2$ > $Cu/CeO_2$ > $Co/CeO_2$ > $Fe/CeO_2$. Our model reproduces this trend except for Cu, for which the TOF is underpredicted. This discrepancy can be explained by, e.g., the effect of active sites not included in our study such as the ones at the metal-support interface, which was shown to have an impact on direct WGS.[63]

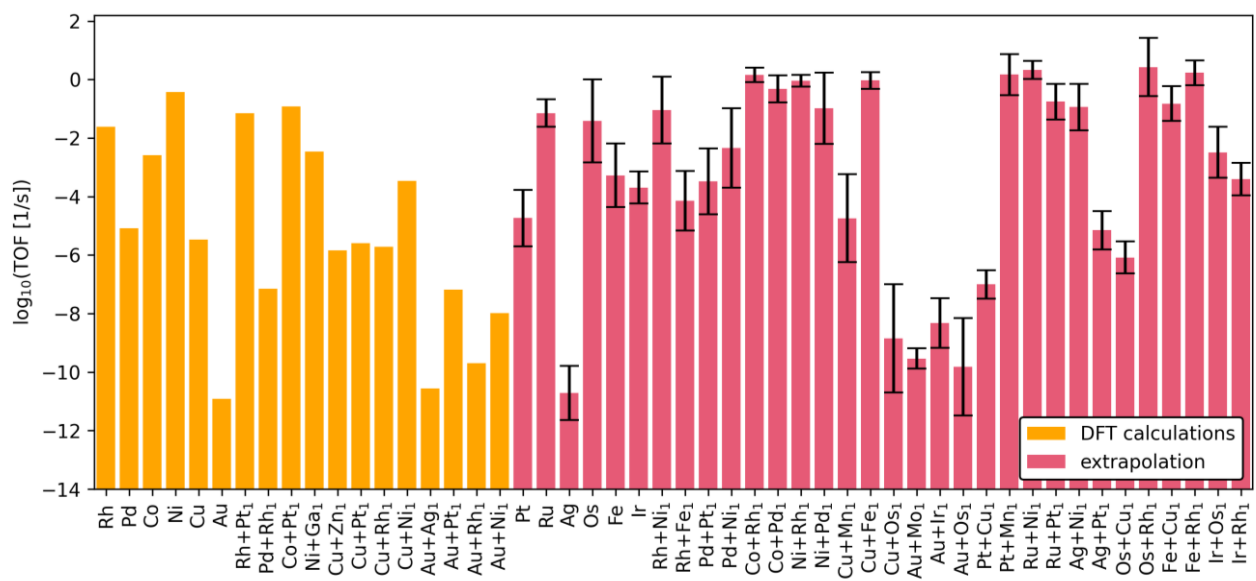

Figure 7: DFT-predicted TOFs of all materials in the train dataset (orange) as well as out-of-domain predictions for further materials (red) based on input from an ensemble of WWL-GPR models (error bars from ensemble standard deviation).

Oxophilic metals like Ni and Rh show high $CO_2$ conversion rates, in agreement with previous reports.[64] However, these metals typically suffer from low CO selectivity due to the competing methanation reaction and coke formation (especially for Ni). Previous studies show that doping Ni can enhance the selectivity towards RWGS.[65–69] Our results show good performances of Ni and Rh SAA (e.g., Ni+$Rh_1$, Ni+$Pd_1$, Rh+$Pt_1$, Rh+$Ni_1$), which makes them possible good candidate catalysts, but further studies on the competing methanation pathways should confirm this.

Co, Ru, and Fe show high O* coverage (see Figure S4), which is detrimental to their catalytic performance by blocking active sites. This observation aligns with the findings of Qi et al.[11], who reported that these metals can exhibit reduced activity due to excessive O* adsorption. However, when doped with other metals (e.g., Co+$Pt_1$, Co+$Rh_1$, Co+$Pd_1$, Ru+$Ni_1$, Ru+$Pt_1$, Fe+$Cu_1$, Fe+$Rh_1$), their predicted TOF improves, which can be attributed to the decreased O* coverage (see Figure S4). Note that for simplicity we here model all materials in the fcc crystal structure, which may alter the adsorption energies compared to, e.g., the bcc crystal structure. Furthermore, it is important to note that these catalysts, also when doped, can potentially transform into oxides under experimental conditions, which could alter their catalytic behavior.

However, the model still provides valuable insights into the intrinsic activity trends and can serve as a predictive tool for further optimization of catalyst materials.

Coinage metals such as Cu, Au, and Ag exhibit lower TOF for the RWGS reaction. In contrast, certain SAAs based on Cu and Ag – e.g., Cu+Ni$_1$, Cu+Fe$_1$, Ag+Ni$_1$ – demonstrate significantly higher catalytic activity. This observation aligns with previous studies showing that doping Cu can enhance its performance as a RWGS catalyst.[11,67,70] In particular, CuFe catalysts showed promising potential for $CO_2$ hydrogenation reactions.[71–73] This is a particularly attractive finding, as Cu and Ag are earth-abundant and inexpensive metals, offering a more economically viable alternative to traditional noble metal catalysts.

For most materials, the CO-O dissociation pathway involving direct $CO_2$* dissociation is dominant (see Figure S5), in agreement with previous findings.[60,61,74] The COO-H pathway is preferred instead by Au, most Au-based single-atom alloys (SAAs), and the (111) surfaces of Pd, Pt, and Ag, as well as by Pt+Cu$_1$ and Ag+Pt$_1$. The H-COO pathway is only relevant for Co(111). $CO_2$* dissociation is the rate-determining step (RDS) for the majority of systems (see Figure S6), while in those favoring the COO-H pathway, the RDS shifts to $CO_2$* hydrogenation to COOH*. On Au(111), the dissociation of COOH* into CO* and OH* becomes the limiting step. Hydrogenation of $CO_2$* to HCOO* is only relevant for Co(111). In addition, $H_2O$* formation is the RDS for a broad range of materials (e.g., Ru, Os, Co(100), Fe(100), Co(100)+Rh$_1$, Ni(100)+Rh$_1$, and other SAAs), highlighting the advantage of using SAAs over metal particles, as $H_2O$* formation breaks the BEP relationship on SAA surfaces (see Figure 3.c).

In summary, our results highlight the potential of SAA materials as promising catalysts for the RWGS reaction. For the case of Cu and Ag, showing low RWGS catalytic activity, the doping with more oxophilic elements such as Ni and Fe significantly enhances the catalytic performances. On the other hand, doping Co, Ru and Fe can mitigate excessive O* coverage that would otherwise block the active sites. Additionally, certain SAAs provide lower TS energies than the BEP relationship of pure metals for the kinetically relevant $H_2O$* formation step. The WWL-GPR, providing accurate adsorbate and TS energy predictions, helps to get a detailed picture of how dopants modulate adsorption energies, reaction pathways, and rate-determining steps. These findings offer a compelling route toward the design of efficient and economically viable RWGS catalysts. Further studies including support effects and selectivity analysis will be essential to fully assess the practical applicability of these materials.

## Conclusions

In this study, we demonstrated the potential of combining graph-based machine learning (ML) models with density-functional theory (DFT) and microkinetic modeling for the efficient screening of catalysts for the reverse water-gas shift (RWGS) reaction on metals and single-atom alloy (SAA) catalysts. By extending the Wasserstein Weisfeiler-Lehman Gaussian Process Regression (WWL-GPR) model to predict both adsorption and transition state (TS) energies, we significantly improved the accuracy of adsorbate and TS energy predictions compared to traditional linear scaling relations. Our results show that using energies calculated with WWL-GPR reduces the errors associated with turnover frequency (TOF) predicted with microkinetic modeling by nearly an order of magnitude compared to conventional linear methods. Furthermore, the application of the proposed methodology to screen a diverse set of metals and SAA catalysts revealed several promising candidate materials for RWGS. Overall, this work not only advances the state of catalyst screening for RWGS but also establishes a robust framework for future studies aiming to design catalysts for other complex surface reactions. The combination of DFT, advanced ML techniques, and microkinetic modeling holds great potential for accelerating the discovery of high-performance materials for energy conversion and sustainable chemical processes.

## Data availability

The DFT structures employed in this work will be uploaded to the NOMAD Repository after publication.

## Supporting information

The Supporting Information is available free of charge and contains a table of the features used in the ML models, additional parity plots comparing DFT and ML predictions, facets fractions obtained from Wulff constructions and coverage, reaction path and degree of rate control analysis from the microkinetic models.

## Acknowledgments

M.A. acknowledges funding from the Aarhus University Research Foundation, the Danish National Research Foundation through the Center of Excellence 'InterCat' (grant no. DNRF150), and VILLUM FONDEN (grant no. 37381). R.C. and M.A. acknowledge support from European Union's Horizon Europe research and innovation programme under the Marie Skłodowska-Curie grant no. 101108769 and 754513, respectively. Computational support was provided by the Centre for Scientific Computing Aarhus (CSCAA) at Aarhus University.

**Table of Contents**

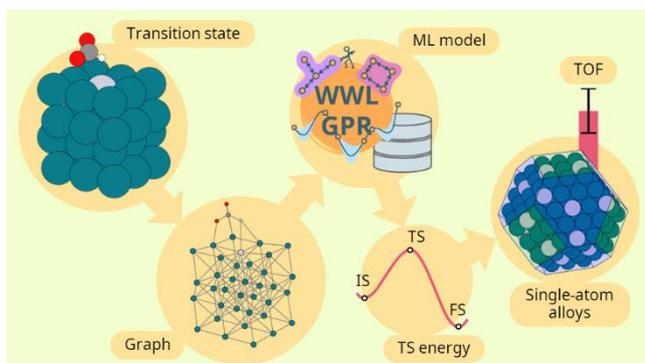

# Supporting Information

# Transition States Energies from Machine Learning: An Application to Reverse Water-Gas Shift on Single-Atom Alloys

Raffaele Cheula[1] and Mie Andersen[1,2]*

[1]Center for Interstellar Catalysis, Department of Physics and Astronomy, Aarhus University, 8000 Aarhus C, Denmark

[2]Aarhus Institute of Advanced Studies, Aarhus University, 8000 Aarhus C, Denmark

*mie@phys.au.dk

Table S1: Features included in this work for the application of the WWL-GPR model. These include features of atomic species (i.e., tabulated data of the chemical elements), features calculated from the PDOS of clean surface, work function of clean surface, features calculated from the adjacency matrix of the target structure, HOMO and LUMO of the adsorbate molecule (in the gas phase), energies of the images (only for TS structures), and SOAP descriptor of clean surface and adsorbate molecule. In the application of EDTs, we use the same features used in the WWL-GPR model but averaged over the atoms of the surface site, i.e., atoms bonded to the adsorbate (or TS).

| | | |
|---|---|---|
| (1) | Electron affinity ($E_{affinity}$) | atomic species |
| (2) | Pauling electronegativity ($EN_{Pauling}$) | atomic species |
| (3) | Ionization potential ($I_{potential}$) | atomic species |
| (4) | Radius of d orbitals ($d_{radius}$) | atomic species |
| (5) | d-band filling ($d_{filling}$) | surface PDOS |
| (6) | d-band center ($d_{center}$) | surface PDOS |
| (7) | d-band width ($d_{width}$) | surface PDOS |
| (8) | d-band skewness ($d_{skewness}$) | surface PDOS |
| (9) | d-band kurtosis ($d_{kurtosis}$) | surface PDOS |
| (10) | sp-band filling ($sp_{filling}$) | surface PDOS |
| (11) | Density of d states at Fermi level ($d_{density}$) | surface PDOS |
| (12) | Density of sp states at Fermi level ($sp_{density}$) | surface PDOS |
| (13) | Surface workfunction ($W_{function}$) | surface |
| (14) | Coordination number ($n_{coord}$) | adjacency matrix |
| (15) | Number of bonds between adsorbate (or TS) and surface ($n_{bonds}$) | adjacency matrix |
| (16) | Highest Occupied Molecular Orbital of gas molecule (HOMO) | molecule |
| (17) | Lowest Unoccupied Molecular Orbital of gas molecule (LUMO) | molecule |
| (18) | Energy of initial state ($E_{first}$) | images |
| (19) | Energy of final state ($E_{last}$) | images |
| (20) | Reaction enthalpy ($\Delta E_R$) | images |
| (21–90) | SOAP descriptor of clean surface and gas phase (SOAP) | molecule & surface |

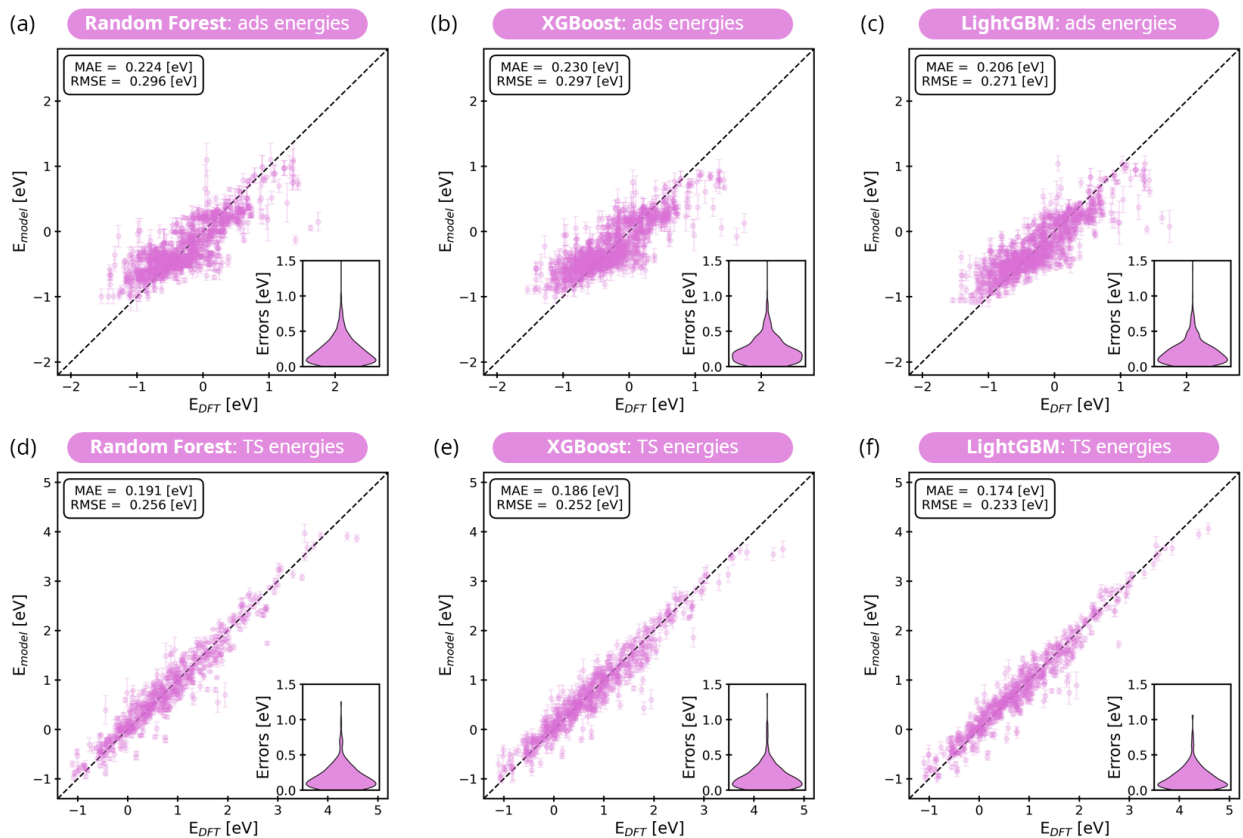

Figure S1. Parity plots comparing DFT-calculated adsorption (ads) and transition state (TS) energies with predictions from different ensemble decision tree models: (a, d) Random Forest, (b, e) XGBoost, and (c, f) LightGBM. Insets show the error distributions.

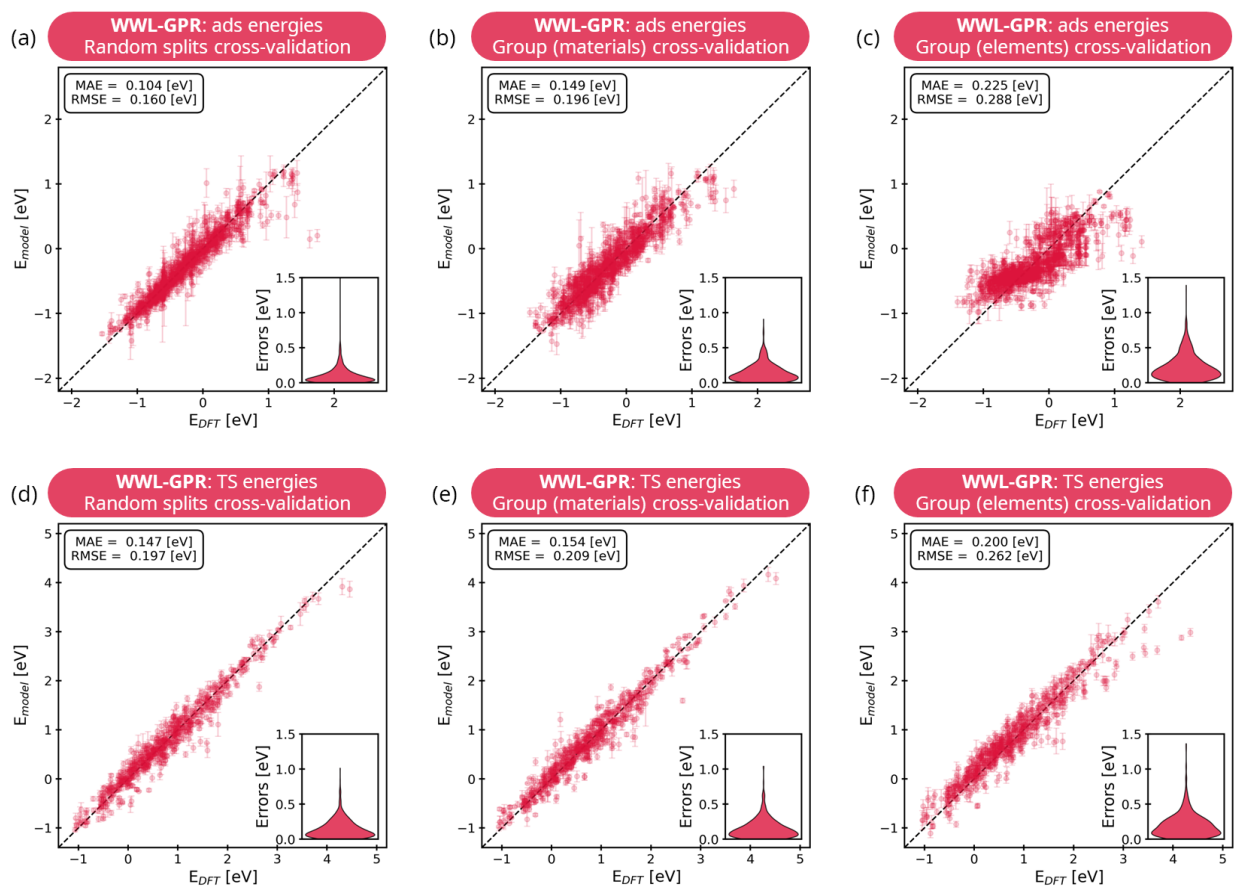

Figure S2. Parity plots comparing DFT-calculated adsorption (ads) and transition state (TS) energies with predictions from the WWL-GPR model, with data split with different cross-validation: (a, d) stratified cross-validation, (b, e) stratified group (materials) cross-validation, and (c, f) stratified group (elements) cross-validation. Insets show the error distributions.

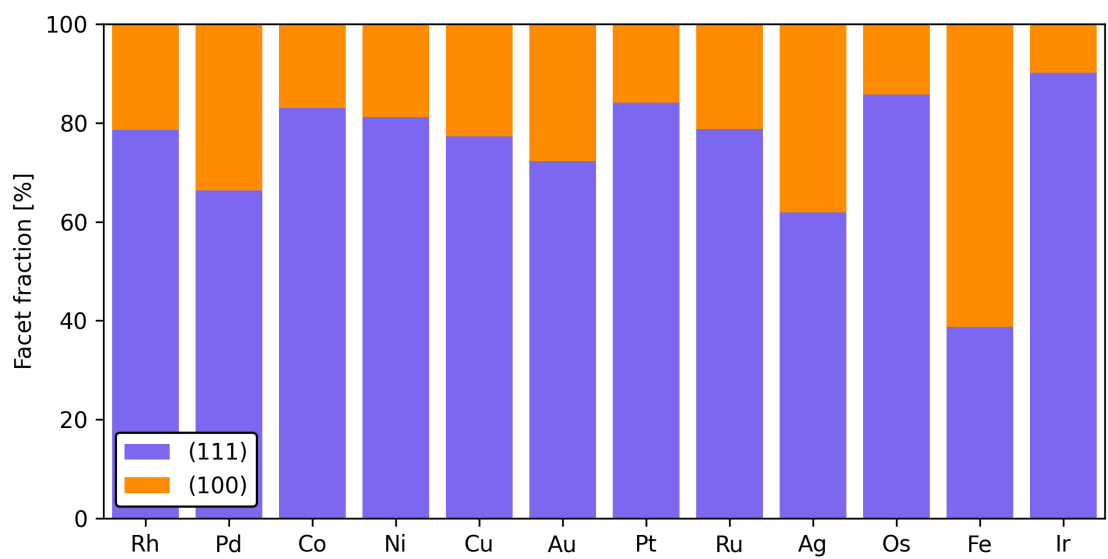

Figure S3: Facets fractions obtained with a Wulff construction of the pure metals, considering the (100) and (111) facets.

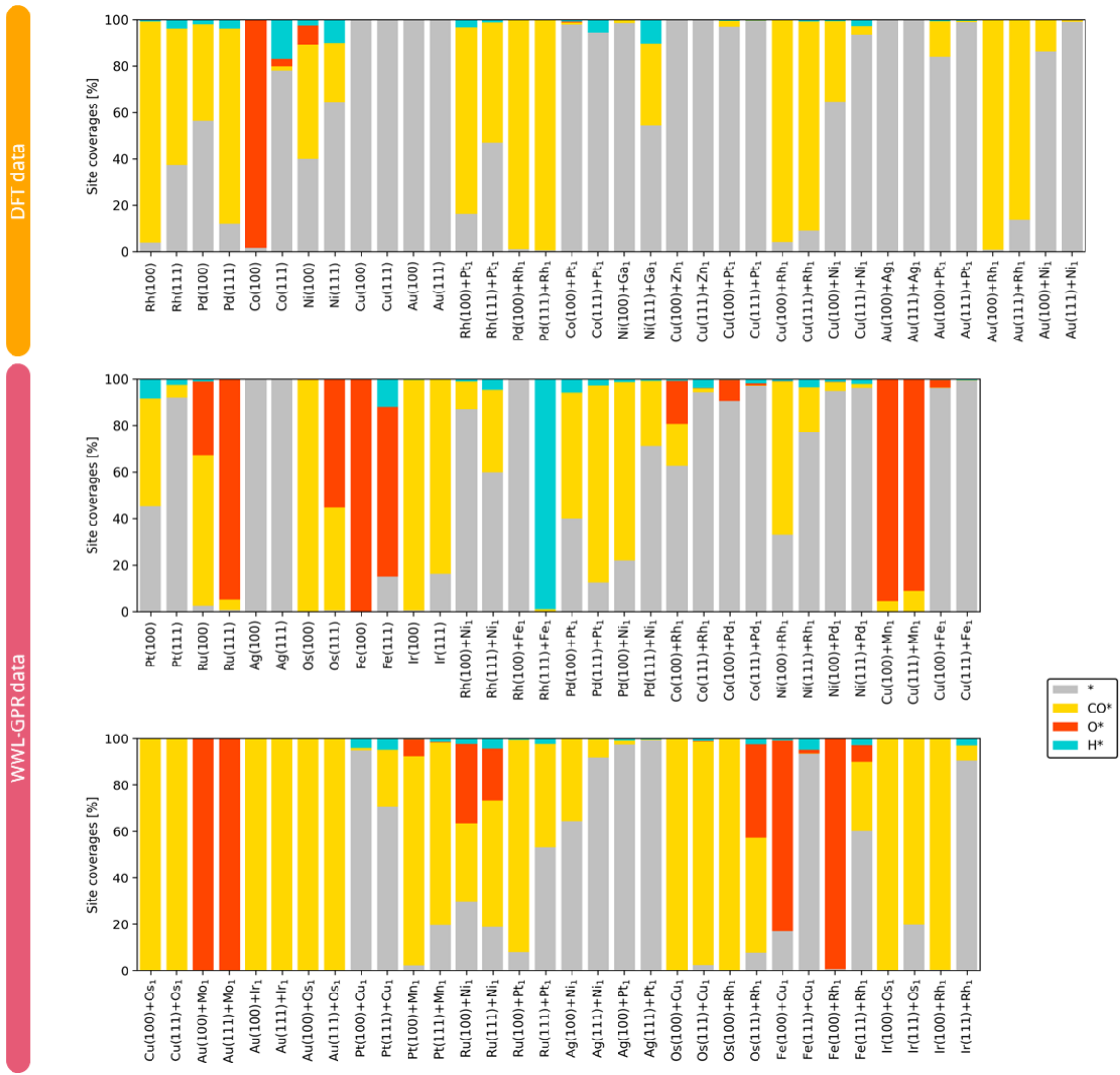

Figure S4. Coverages of reaction intermediates during RWGS on different surfaces, calculated with microkinetic models based on energy data from DFT (top row) and WWL-GPR (bottom two rows).

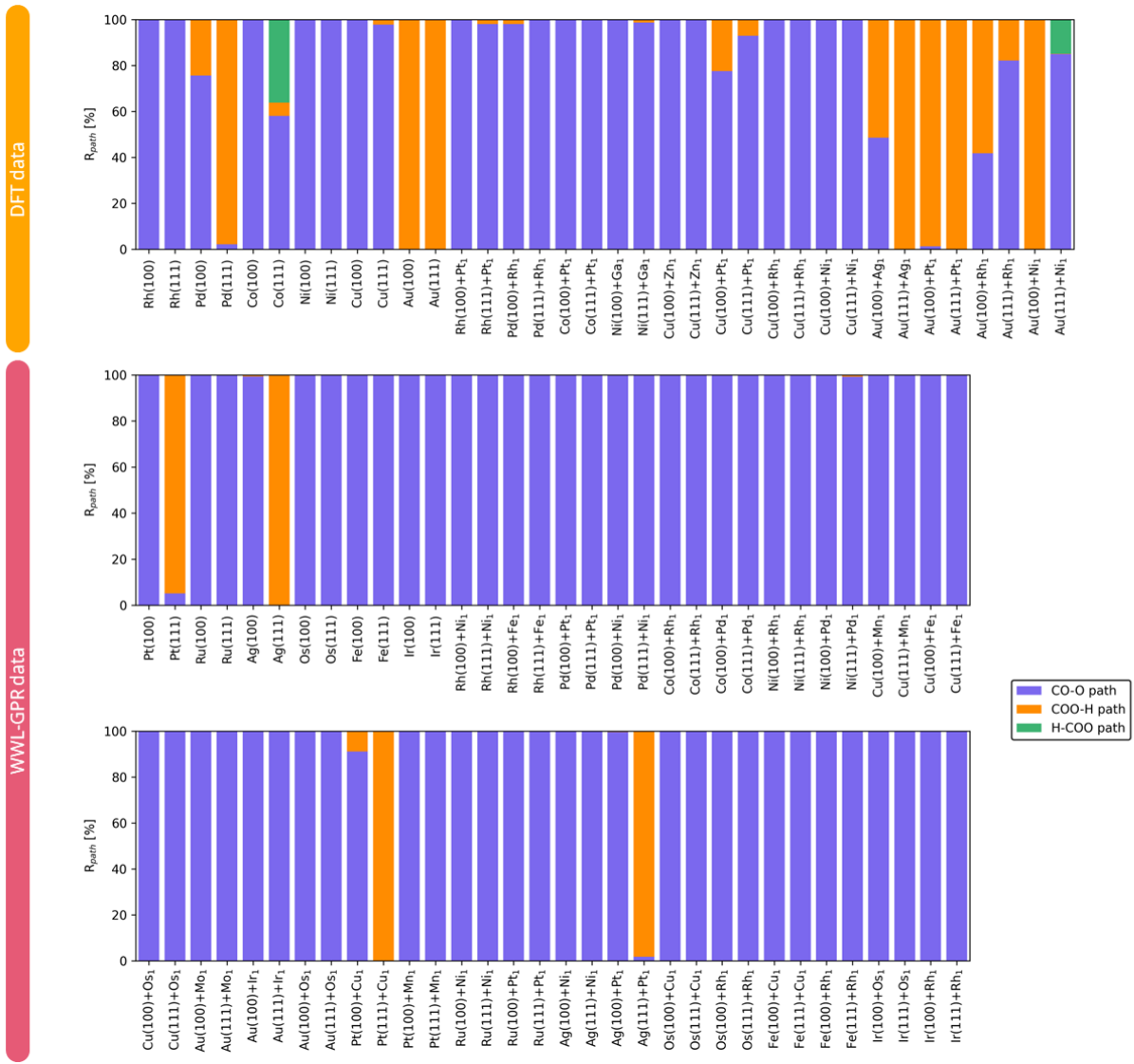

Figure S5. Reaction paths of RWGS on different surfaces, calculated with microkinetic models based on energy data from DFT (top row) and WWL-GPR (bottom two rows).

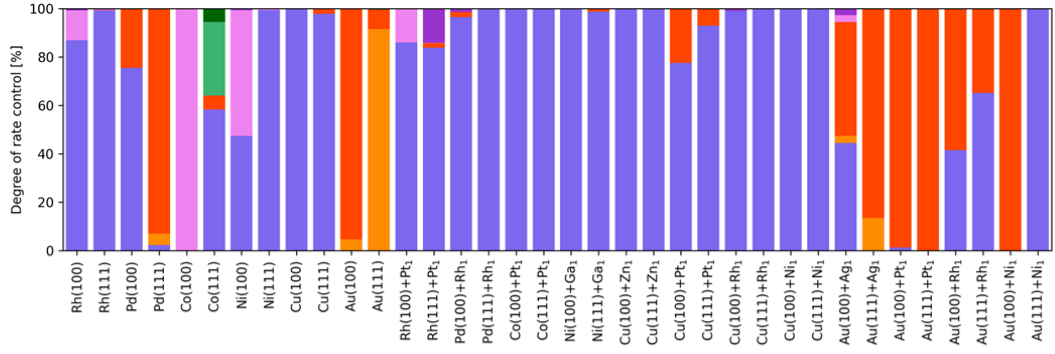
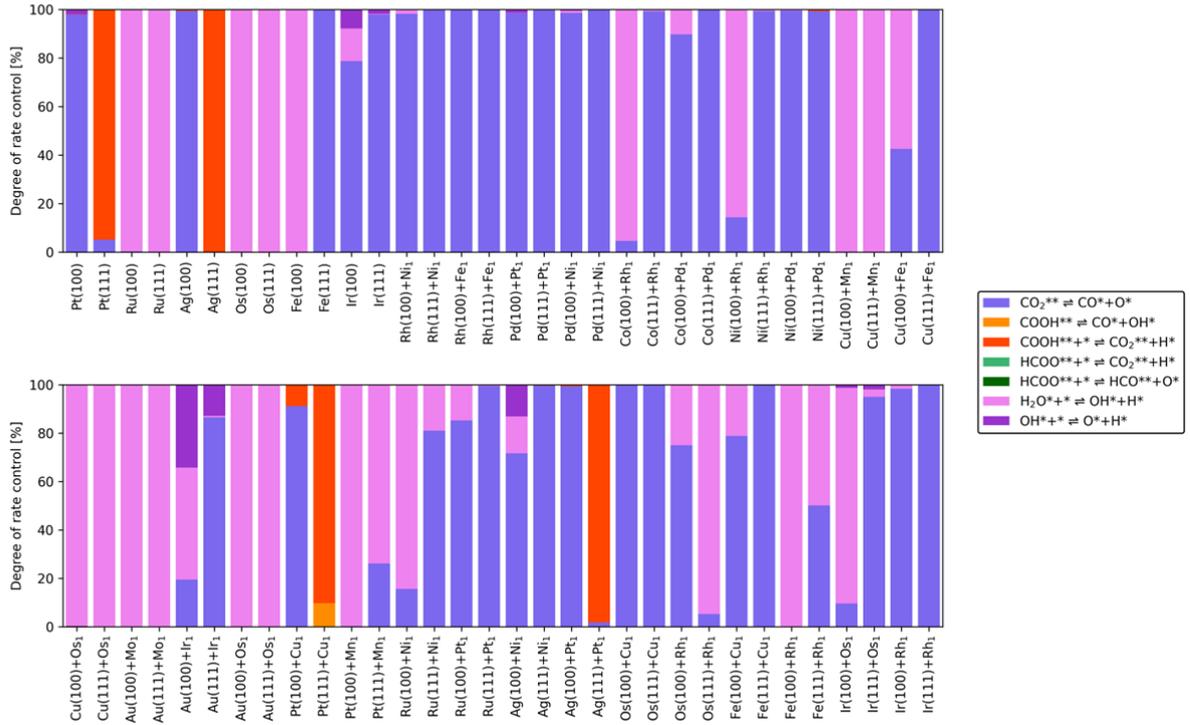

Figure S6. Degrees of rate control of the elementary steps of RWGS on different surfaces, calculated with microkinetic models based on energy data from DFT (top row) and WWL-GPR (bottom two rows).